\documentclass[a4paper,twoside]{article}

\usepackage{epsfig}
\usepackage{subcaption}
\usepackage{calc}
\usepackage{amssymb}
\usepackage{amstext}
\usepackage{amsmath}
\usepackage{amsthm}
\usepackage{multicol}
\usepackage{pslatex}
\usepackage{apalike}
\usepackage{algorithm2e}
\usepackage[bottom]{footmisc}
\usepackage{caption}
\usepackage{amsmath}
\usepackage{amssymb}
\usepackage{placeins}

\usepackage{braket}

\usepackage{SCITEPRESS}     

\usepackage{tikz}
\usetikzlibrary{quantikz2}

\newcommand{\rom}[1]{\uppercase\expandafter{\romannumeral #1\relax}}

\begin{document}

\title{Variational Quantum Circuit Design for Quantum Reinforcement Learning on Continuous Environments}

\author{\authorname{Georg Kruse\sup{1}, Theodora-Augustina Dr{\u{a}}gan\sup{2}, Robert Wille\sup{3} and Jeanette Miriam Lorenz\sup{4}}
\affiliation{\sup{1}Fraunhofer Institute for Integrated Systems and Device Technology, Erlangen, Germany}
\affiliation{\sup{1,3}Technical University of Munich, Department of Informatics, Munich, Germany}
\affiliation{\sup{2,4}Fraunhofer Institute for Cognitive Systems IKS, Munich, Germany}
\affiliation{\sup{4}Ludwig Maximilian University, Faculty of Physics, Munich, Germany}
\email{\sup{1}georg.kruse@iisb.fraunhofer.de, \sup{2}theodora-augustina.dragan@iks.fraunhofer.de, \sup{3}robert.wille@tum.de, \sup{4}jeanette.miriam.lorenz@iks.fraunhofer.de}
}


\abstract{Quantum Reinforcement Learning (QRL) emerged as a branch of reinforcement learning (RL) that uses quantum submodules in the architecture of the algorithm. One branch of QRL focuses on the replacement of neural networks (NN) by variational quantum circuits (VQC) as function approximators. Initial works have shown promising results on classical environments with discrete action spaces, but many of the proposed architectural design choices of the VQC lack a detailed investigation. Hence, in this work we investigate the impact of VQC design choices such as angle embedding, encoding block architecture and postprocessesing on the  training capabilities of QRL agents. We show that VQC design greatly influences training performance and heuristically derive enhancements for the analyzed components. Additionally, we show how to design a QRL agent in order to solve classical environments with continuous action spaces and benchmark our agents against classical feed-forward NNs.}

\onecolumn \maketitle \normalsize \setcounter{footnote}{0} \vfill

\section{\uppercase{Introduction}}
\label{sec:introduction}

Quantum computing (QC) is a research field that is drawing a lot of attention due to the expected computational advantages. There are many possible application fields, such as quantum chemistry, cryptography, search algorithms and others~\cite{dalzell2023quantum}. Moreover, quantum hardware is becoming increasingly accessible, with noisy intermediate scale quantum (NISQ) devices already being available. This creates the possibility of designing, implementing and benchmarking QC algorithms that are NISQ-friendly and comparing them against classical methods in order to assess potential quantum advantage already at the current state of technology.

Quantum machine learning is one of the most promising candidates to show quantum advantage on NISQ hardware. Variational quantum algorithms (VQA) for supervised learning~\cite{Perez_Salinas_2020}, for unsupervised learning~\cite{Benedetti_2019,yuxuan2020expressive}, and for reinforcement learning~\cite{jerbi2021parametrized,skolik2022quantumGym} have been proposed and have already been implemented on NISQ machines. In supervised learning, neural networks (NN) were replaced with variational quantum circuits (VQC). While initial studies suggest that VQCs inhibit preferable properties such as better trainability~\cite{McClean2018Barren}, other analyses of important properties such as learning capability and generalization errors \cite{abbas2021power,caro2022generalization,banchi2021generalization} remain inconclusive with regard to the advantages of quantum computation. Whether VQCs show reliable advantage over NNs therefore remains an open question~\cite{qian2022dilemma}.

Reinforcement learning (RL) is a paradigm of machine learning where an agent learns by interacting with the environment. Current literature in RL proposes agents that are able to tackle important and complex problems in the field of energy management~\cite{yu2021review}, healthcare~\cite{yu2020reinforcement}, and robotics~\cite{haarnoja2019soft}. Since generalization~\cite{pmlr-v97-cobbe19a} as well as sample efficiency are potential advantages of QC and important aspects of RL, it is therefore of interest to investigate potential enhancements QC could bring to RL. 

The literature in the subdomain of quantum reinforcement learning (QRL) is yet sparse. Multiple approaches have been proposed and can be divided into several categories, ranging from quantum-inspired methods that mainly use classical computation, to purely quantum approaches that require fault-tolerant devices that are not yet available~\cite{meyer2022survey}. A main branch of research are hybrid quantum-classical algorithms that contain VQCs as function approximators whose trainable parameters are updated using classical methods, such as gradient descent. This branch of QRL, also referred to as VQC-based QRL, is of special interest since the possible beneficial properties of VQCs such as better trainability and generalization \cite{abbas2021power,banchi2021generalization} can be transferred to RL algorithms. In this branch of research, quantum advantage has already been shown on an artificial benchmark \cite{jerbi2021parametrized}. 

In classical RL, as well as in NN training in general, hyperparameters greatly influence the performance while strongly varying across different tasks, making an a priori design choice difficult~\cite{smith2018disciplined}. A plethora of heuristics has therefore been developed by mainly empirical methods in order to find suitable architectures and proper hyperparameters for a given task. This intensive numerical study for classical NNs remains to be conducted for VQCs. Recent works have mainly followed the architecture and hyperparameter choices of previous publications~\cite{jerbi2021parametrized,skolik2022quantumGym}, albeit these choices have been insufficiently investigated. 

The design of VQCs remains a difficult task, since many VQC architectures suffer from various drawbacks, such as barren plateaus~\cite{McClean2018Barren} and trainability issues~\cite{zhang2020trainability}. The latter was also studied empirically and theoretically by Anschuetz et al.~\cite{anschuetz2022quantum}, who argue that a large class of multi-layer VQCs are untrainable. Similarily, Larocca et al. have shown that -- just like for NNs -- various VQCs have to be overparameterized in order to be trainable~\cite{larocca2023theory}, hindering the use of VQCs with larger qubit numbers, which are necessary to tackle more complex tasks. While Skolik et al. proposed VQCs which already contain major advances such as data reuploading and trainable output weights, other design choices like the data encoding method, which is just as important as general VQC architectural design choices~\cite{schuld2021effect}, have barely been explored. VQC design for QRL remains thus an open question.

While the majority of QRL literature focuses on algorithms for environments with discrete action spaces, Wu et al. proposed a QRL solution for \textit{quantum} continuous action space (CAS) environments~\cite{wu2020quantum}. While Lan and Acuto et al. model QRL agents on \textit{classical} CAS environments~\cite{lan2021variational,acuto2022variational}, they still use additional NNs as post-processing layers. This approach makes it difficult to distinguish between the contribution of quantum and classical part of the algorithm. Another open question for VQC-based QRL is therefore the adaptation to CAS environments without the use of additional classical NNs.

Based on the identified gaps in literature on the construction of QRL algorithms and the design of VQCs, our contributions are as follows: First, we show how to design VQC-based QRL for classical CAS environments without the use of additional NN as pre- or postprocessing layers. Second, we investigate VQC design choices by analysing the influence of angle embedding, encoding block design and readout strategies on the performance of the agent, benchmarked against two classical CAS OpenAI Gym environments~\cite{brockman2016openai}, Pendulum-v1 and LunarLander-v2.

\begin{figure*}
    \centering
    \includegraphics[width=0.95\textwidth]{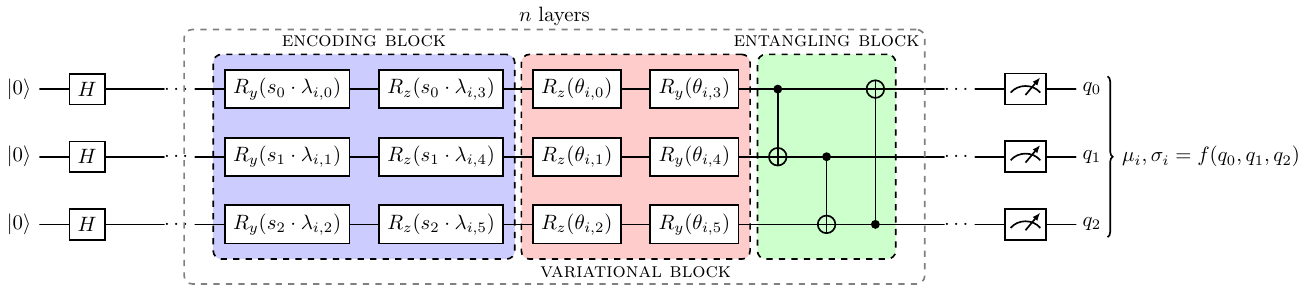}
    \caption{An exemplary VQC with three qubits consists of multiple layers $n$. Each layer has three blocks: an encoding block (with input state $s$ and scaling parameters $\lambda$), a variational block (with variational parameters $\theta$) and an entangling block (a daisy chain of CNOT entangling gates), followed by measurement and postprocessing steps.}
    \label{fig:vqc_architecture}
\end{figure*}

\section{\uppercase{Quantum Preliminaries}}
In classical computing, one uses bits and strings of bits to encode information in one of two possible states 0 or 1, whereas in quantum computing the basic unit of information is the quantum bit -- or, for short, the qubit. A qubit, opposed to a bit, can be in an infinite amount of states and is usually represented as a superposition of two basis states $\ket{0}$ and $\ket{1}$ in the following manner:
\begin{equation}
    \psi = \alpha \ket{0} + \beta \ket{1},
\end{equation}
where $\alpha, \beta \in \mathbb{C}$. When the qubit is measured, it collapses to one of the basis states with probabilities $p_0 = |\alpha|^2$ and $p_1 = |\beta|^2$, respectively. While classical bit values are stored and processed using electrical signals and circuits, qubits are initialised, stored and manipulated using physical systems that are able to simulate or are themselves quantum systems. Qubits are manipulated by quantum gates, which are operations that act on one or multiple qubits and transform their state, changing their probabilities. A series of multiple gates is called a quantum circuit, and if some parameters of these gates are trainable, it becomes a VQC.

When it comes to quantum computation, either or both the input data and the algorithm can be classical or quantum. In this work, we focus on the subfield of QRL where the data is classical and the algorithm uses a hybrid quantum-classical approach, which contains VQCs as function approximators. The general architecture of a VQC used in this work is depicted in Fig.  \ref{fig:vqc_architecture}. It consists of three qubits, represented by three horizontal lines, which are initialized in the basis state $\ket{000}$. On these qubits act a sequence of quantum gates, indicated by the boxes on these lines, which change the state of the qubits. The gates are separated into three different blocks: A data encoding block, which transforms qubits depending on the classical input, a variational block with trainable variational gates, and an entangling block, where two qubit gates are used to entangle the qubits. Together the three blocks form a layer, which can be repeated several times. The repetition of a data encoding block in a VQC is known as data reuploading. At the end of the VQC, the qubits are measured and, if necessary, a classical post-processing step is applied to adapt the output of the measurement to the task at hand.

Quantum circuits, such as the aforementioned VQC, can be executed either in a classical simulation or on real quantum hardware. The quantum devices available today are NISQ devices, whose physical realizations are still affected by noise and offer the user a limited number of qubits as well as qubit operations, which is in contrast to the ideal assumptions of fault-tolerant algorithms. This also motivates the approach and design choices in this work, as one of the current goals is to create shallow NISQ-suitable circuits for VQC-based QRL solutions, which can be run on current quantum technology. In the future, this would allow to benchmark the performance of the algorithm described in this work on real quantum technology as well as to analyse the real potential quantum advantage.

\section{\uppercase{Related Works}}

In the branch of QRL this work focuses on, classical RL algorithms are modified by replacing parts of the computational process with VQCs. Among these RL algorithms one can find the deep Q-learning algorithm~\cite{skolik2022quantumGym}, as well as Actor-Critic (AC) methods: Proximal Policy Optimization (PPO)~\cite{druagan2022quantum}, the Soft Actor-Critic (SAC)~\cite{acuto2022variational}, as well as the Asynchronous Advantage Actor-Critic (A3C)~\cite{chen2023asynchronous}. These solutions solve both classical and quantum environments.

A hybrid quantum-classical deep Q-learning QRL algorithm, that integrates VQCs as function approximators, is proposed in~\cite{skolik2022quantumGym} and benchmarked on two classical Gym environments: the non-slippery stochastic Frozen Lake and the Cart-Pole. The general structure of the VQC is of repeated layers of rotational gates for both the angle embedding of input data and for the trainable blocks of the model, followed by a circular entanglement with CZ gates of the qubits. The authors mention how the algorithm would be adapted to continuous actions, namely by discretizing the action into bins, but did not detail the method or benchmark it on CAS environments. Moreover, encoding and measurement strategies are described and the importance of the choice of these components is mentioned, but without a detailed analysis of the influence on the performance. 

Many recent publications have afterwards adopted these design choices~\cite{sun2023differentiable,chen2023asynchronous}. For example, in~\cite{sun2023differentiable} a framework is proposed to learn the most suitable architecture for quantum deep Q-learning on the Frozen Lake and on the Cart-Pole environments. However, the search space of the circuit architectures is still restricted to the models described in~\cite{skolik2022quantumGym}.

The author of~\cite{chen2023asynchronous} proposes hybrid classical-quantum A3C models that successfully solve three Gym environments~\cite{brockman2016openai}, namely the Acrobot, the Cart-Pole and the MiniGrid~\cite{chevalier2018minimalistic}. While the input vectors are continuous, the action space is still discrete, with the agent being able to choose between at most six different actions to perform. Before and after the VQC a single NN linear layer is used for data pre- and post-processing. The former is to convert the input dimensions in order to fit the eight-qubit architecture, while the latter is to adapt the output to the format and size dictated by the environment characteristics. Their VQC design consists of an angle  embedding block, followed by two trainable layers of CNOT entanglement gates and trainable rotational gates across all three axes. While this VQC-based solution leads to results comparable to, or even surpassing, its classical counterpart, the usage of NNs makes the potential quantum advantage difficult to assess. Similarly, other contributions solved CAS environments such as the Gym Pendulum-v1~\cite{lan2021variational} or the control of a robotic arm~\cite{acuto2022variational} using a quantum version of the SAC algorithm. This NN-enhanced hybrid quantum-classical approach is also successfully used in quantum multi-agent reinforcement learning for a drone communication environment~\cite{park2023quantum}.

The authors of~\cite{skolik2023robustness} introduce QRL models based on Policy Gradient and on Q-learning and apply them to the Cart-Pole and to the Traveling Salesperson Problem, which are both classical environments with discrete action spaces. They study the robustness of these models under the impact of shot noise and coherent and incoherent errors. They apply trainable weights, once to scale each input feature before encoding it with angle embedding into the VQC, and then to scale the output of the Q-values to the values dictated by the characteristics and task of the environment. 
Lastly, some works focus on quantum environments: the authors of~\cite{wu2020quantum} propose a quantum Deep Deterministic Policy Gradient (DDPG) algorithm and apply it to a CAS task, namely the quantum state generation. The solution is benchmarked on one-qubit and two-qubit cases. While the algorithm is successful, it is not presented how to adapt it to a classical environment, i.e., how to embed the data and interpret the measurements, which are the biggest challenges of CAS environments.
\section{\uppercase{Quantum Actor-Critic}}

\begin{figure*}
    \centering
    \includegraphics[width=0.8\textwidth]{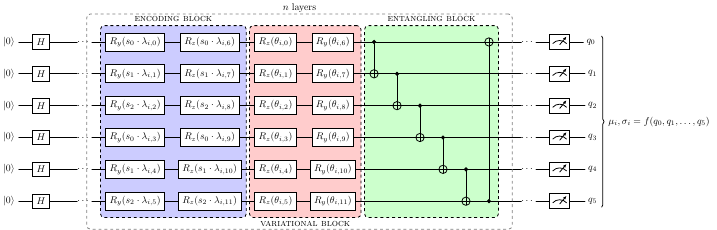}
    \caption{An example of a \textit{stacked} VQC with six qubits applied for an environment with observation space of size three. The encoding block is repeated (stacked) vertically two times such that each state feature $s_i$ is encoded twice on two distinct qubits in each layer $n$. The other blocks follow the design of Fig. \ref{fig:vqc_architecture}.}
    \label{fig:vqc_stacked}
\end{figure*}

As has been shown in previous works, Q-learning \cite{skolik2022quantumGym}, Policy Gradient \cite{jerbi2021parametrized} algorithms, as well as actor-critic algorithms such as PPO \cite{druagan2022quantum}  can be adapted to VQC-based QRL. Building on the works of Dr{\u{a}}gan et al., in this section we show how to advance this approach to CAS environments without the need of additional classical pre- or postprossesing layers. This is especially important, since the class of problems state-of-the-art QRL focuses on is still quite simple. Differentiating between the contribution of quantum and classical part of the algorithm can therefore pose a difficult question. This is why we aim to reduce the complexity of the classical pre- and postprocessing to simple input- and output scalings, rather than entire NN layers as has been previously proposed in QRL solutions for CAS environments~\cite{lan2021variational,acuto2022variational,park2023quantum}.

The PPO algorithm \cite{schulman2017proximal} consists of an actor and a critic, which are each represented by one function approximator (classically a NN). The actor estimates the policy function $\pi_{\Theta}(s_t)$,  while the critic estimates the value function $V_{\Phi}(s_t)$, both at a given state $s_t$ at time step $t$. The gradients of the actor  $\hat{\Theta}_{\text{Act}}$ and the critic $\hat{\Phi}_{\text{Crit}}$ are calculated according to:

\begin{equation}\label{gradients_actor}
\hat{\Theta}_{\text{Act}} = \sum_{t=0}^T \nabla_{\Theta} \log \pi_{\Theta} (a_t | s_t) A^{\pi_{\Theta}}(s_t, a_t),
\end{equation}

\begin{equation}\label{gradients_critic}
\hat{\Phi}_{\text{Crit}} = \sum_{t=0}^T \Big( V_{\Phi}(s_t) - \hat{R_t}\Big)^2,
\end{equation}\smallskip
where $A^{\pi_{\Theta}}(s_t, a_t)$ is the advantage function, $\pi_{\Theta}(a_t|s_t)$ the probability of sampling action $a_t$ in state $s_t$ and $\hat{R_t}$ denotes the discounted reward at time step $t$ (in the following we drop the index $t$ for simplicity).

\subsection{Quantum Actor for Continuous Actions}\label{cont_act}

In order to draw continuous actions from the policy function $\pi_{\Theta}(s)$, the output of the actor needs to be reparameterized. To calculate the value of a given continuous action $a_i$, the function approximator of the actor needs to compute two variables for each action, namely the mean $\mu_i$ and the variance $\sigma_i$ of a normal distribution $\mathcal{N}$ from which the action $a_i$ is then drawn $a_i \sim \mathcal{N}(\mu_i, \sigma_i)$~\cite{schulman2015trust}.

We now consider the computation of the policy $\pi_{\Theta}$ of the actor with a VQC as function approximator instead of a NN. The actor VQC $U_{\Theta}(s)$ is parameterized by input scaling parameters $\lambda_a$, variational parameters $\theta$ and output scaling parameters $w_{\mu_i}$ and $w_{\sigma_i}$, where $\Theta = (\lambda_a, \theta, w_{\mu_i}, w_{\sigma_i})$.
To compute action $a_i$ as a factorized Gaussian, the mean $\mu_i$ and standard deviation $\sigma_i$ are calculated based on the observables $O_{\mu_i}$ and $O_{\sigma_i}$ as follows:
\begin{equation} \label{gaussian_policy_mean}
\mu_i = \bra{0^{\otimes n}} U_{\Theta}(s)^{\dagger} O_{\mu_i} U_{\Theta}(s) \ket{0^{\otimes n}} \cdot w_{\mu_i} 
\end{equation}
and
\begin{equation} \label{gaussian_policy_dev}
\sigma_i = \exp{\big(\bra{0^{\otimes n}} U_{\Theta}(s)^{\dagger} O_{\sigma_i} U_{\Theta}(s) \ket{0^{\otimes n}} \cdot w_{\sigma_i} \big)}.
\end{equation}\smallskip
Since $O_{\mu_i}$ and $O_{\sigma_i}$ are arbitrary Pauli operators, the output values for mean and variance can not scale beyond the interval of [-1, 1]. Therefore the classical scaling parameters $w_{\mu_i}$ and $w_{\sigma_i}$ are crucial in order to apply VQC-based RL to classical CAS environments.

\subsection{Quantum Critic for Value Estimation}

To retrieve the information for the value estimate of the critic, we follow the approach of \cite{skolik2022quantumGym}. Let $U_{\Phi}(s)$ be the critic VQC parameterized by $\Phi = (\lambda_c, \phi, w_{O_{v_i})}$, where analogously to the actor VQC, $\lambda_c$ are the parameters used for input scaling, $\phi$ are the variational parameters, and $w_{O_{v_i}}$ refers to the output scaling parameters. Then the value of a given state $s$ is computed using Eq.~\ref{eq:quantum_critic}
\begin{equation}
    V_{\Phi}(s) = \sum^n  \bra{0^{\otimes n}} U_{\Phi}(s)^{\dagger} O_{v_i} U_{\Phi}(s) \ket{0^{\otimes n}} \cdot w_{O_{v_i}}
    \label{eq:quantum_critic}
\end{equation}\smallskip
We obtain the value of $V_{\Phi}(s)$ by either a single or a sequence of observables $O_{v_i}$ acting on $n$ qubits. We introduce another scaling parameter $w_{O_{v_i}}$, since the value estimate of the critic also needs to scale beyond the interval of $[-1, 1]$ for most RL tasks.
In the following we discuss the choice of the number of VQC layers $n$ and demonstrate how its value, as well as the choice of the observables, can greatly influence QRL performance.

\section{\uppercase{Variational Quantum Circuit Design}}

Due to the small number of empirical studies in the field of QRL, the degrees of freedom in VQC design choices are enormous. In this work we therefore need to restrict our investigations: The basis of our analysis will be the widely-used hardware efficient Ansatz proposed by Jerbi et al. enhanced using data reuploading as proposed by Skolik et al. \cite{jerbi2021parametrized,skolik2022quantumGym}. Our only modification to this Ansatz will be the replacement of CZ entangling gates with CNOT entangling gates. This is due to the fact that chain CZ entanglement may lead to large amounts of parameters which do not influence the  output of the VQC, as the number of qubits increases. A more detailed description of this phenomenon can be found in Fig. \ref{triangular:fig:explanation} in the Appendix.

The basic architecture of the used VQC is depicted in Fig. \ref{fig:vqc_architecture}. Each layer of the VQC consists of three blocks: A data encoding block, a variational block and an entangling block. After $n$ such layers are concatenated, measurements are conducted, followed by an additional postprocessing step. In this work we investigate three design choices for this VQC. First, we evaluate the influence of different preprocessing steps on the classical state $s$ used for angle embedding. Second, we propose a new encoding block architecture and benchmark it against the basic encoding block. Third, we analyse the influence of different observables and postprocessing steps on the training performance.

\begin{figure*}[ht!]
    \centering 
\begin{subfigure}{0.43\textwidth}
  \includegraphics[width=\linewidth]{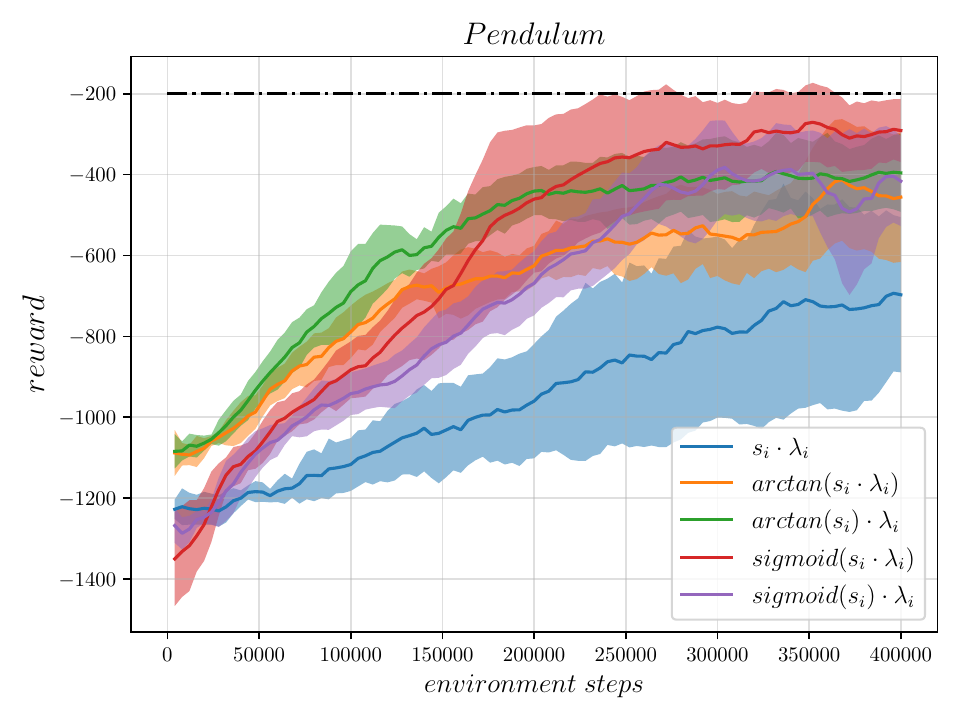}
  \caption{3 qubits, 3 layers, $M_1$ readout, no normalization.}
  \label{pendulum_without_norm}
\end{subfigure}\hfil 
\begin{subfigure}{0.43\textwidth}
  \includegraphics[width=\linewidth]{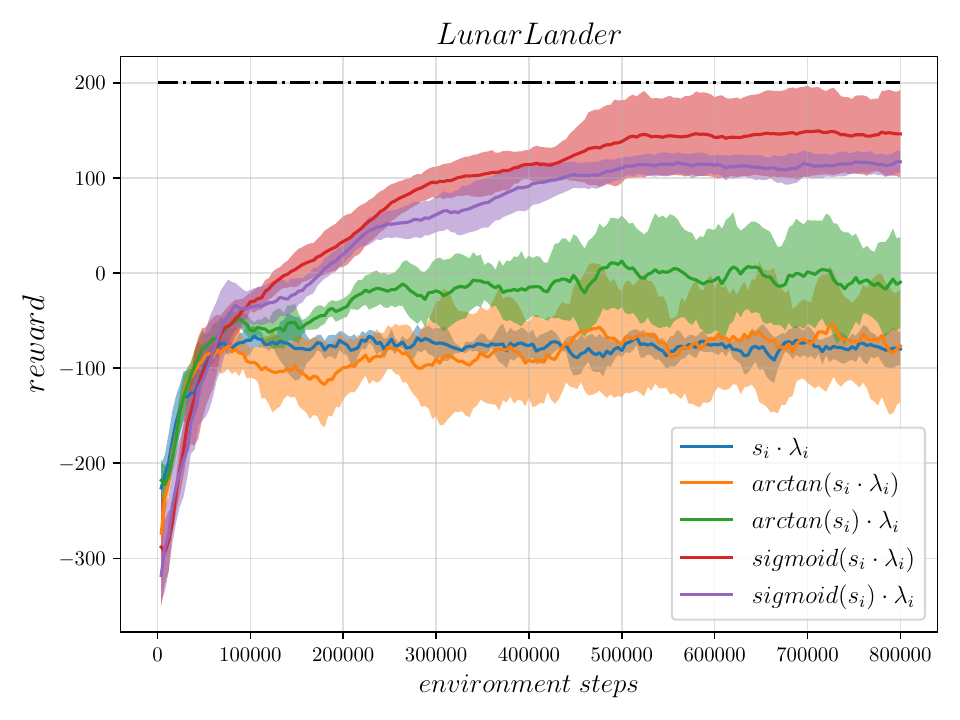}
  \caption{6 qubits, 6 layers, $M_1$ readout, no normalization.}
  \label{lunarlander_without_norm}
\end{subfigure}\hfil
\medskip \vfil
\begin{subfigure}{0.43\textwidth}
  \includegraphics[width=\linewidth]{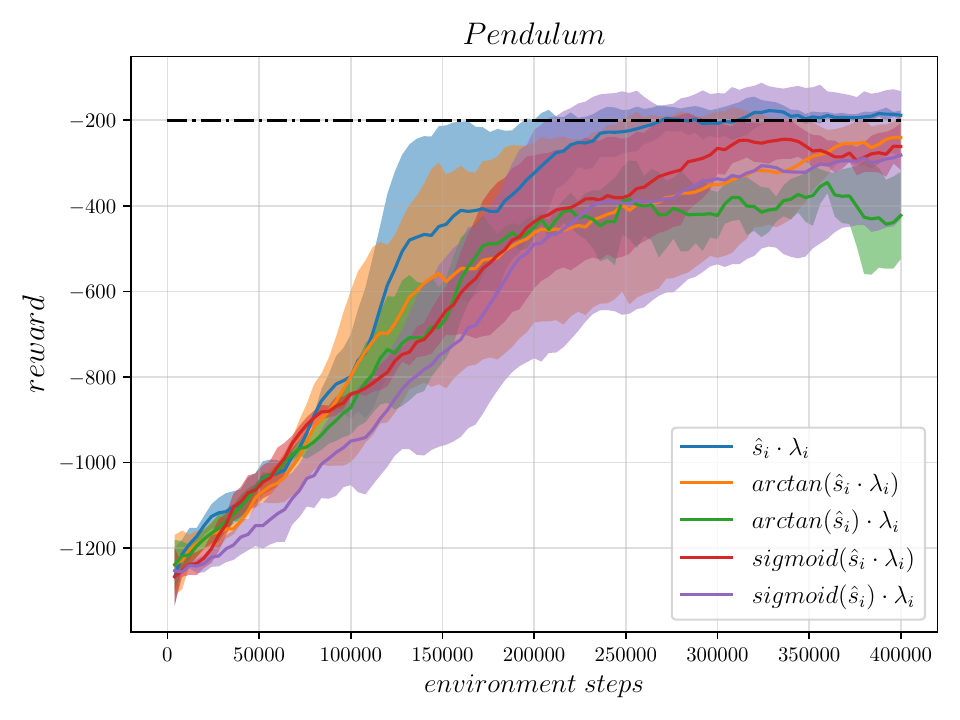}
  \caption{3 qubits, 3 layers, $M_1$ readout, with normalization.}
  \label{pendulum_with_norm}
\end{subfigure}\hfil 
\begin{subfigure}{0.43\textwidth}
  \includegraphics[width=\linewidth]{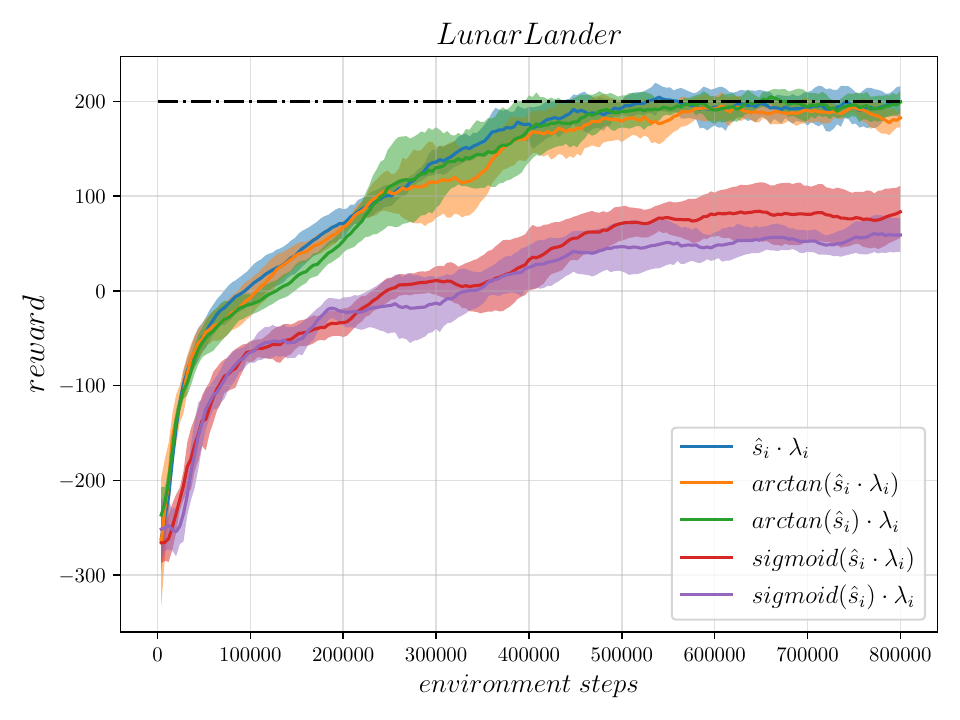}
  \caption{6 qubits, 6 layers, $M_1$ readout, with normalization.}
  \label{lunarlander_with_norm}
\end{subfigure}\hfil \vspace{0.5em}
\caption{Evaluation of encoding strategies: Each angle embedding consists of rotation gates $R_y$ and $R_z$ (ref. Fig. \ref{fig:vqc_architecture}) and the same readout $M_1$ (compared in Fig. \ref{fig_measurement}) is used for all runs. The target reward of the environments are indicated by the dotted black line. In Fig. \ref{pendulum_without_norm} and \ref{lunarlander_without_norm} the training curves for Pendulum-v1 and LunarLander-v2 are depicted without a previous normalization of the state $s$. In Fig. \ref{pendulum_with_norm} and \ref{lunarlander_with_norm} the state $s$ is previously normalized to an interval of $[-\frac{\pi}{2}, \frac{\pi}{2}]$. Each solid line represents the mean of five seeds, the shaded area indicates the standard deviation.}
\label{fig_encoding}
\vfil
\end{figure*}

\subsection{Angle Embedding}\label{angular_encoding}

Data encoding greatly influences the behaviour of VQCs \cite{schuld2021effect}. One of the ways the classical environment state $s$ can be encoded into a quantum state suitable for a VQC is angle embedding. This is done using one or more rotation gates. In this work we follow the approach of~\cite{jerbi2021parametrized} with two rotation gates $R_y$ and $R_z$ (ref. Fig. \ref{fig:vqc_architecture}). Since these gates have a periodicity of $2\pi$ while the observation space of a classical environment can be outside this interval, various works have proposed to encode each feature $s_i$ of the classical environment state $s$ as  $\arctan(s_i \cdot \lambda_i)$ into the two rotation gates ~\cite{skolik2022quantumGym}, where $\lambda_i$ denotes a classical trainable scaling parameter. This encoding has the caveat that for classical observation spaces with large absolute feature values, trigonometric transformations such as $arctan$ and $sigmoid$ will make the features almost indistinguishable for the QRL agent. To overcome this caveat, we propose to previously normalize the features $s_i$ to an interval of $[-\frac{\pi}{2}, \frac{\pi}{2}]$. The normalized features $\hat{s_i}$ are encoded into the rotation gates as  $\hat{s_i}\cdot\lambda_i$, either with or without an additional nonlinear transformation (ref. Fig. \ref{fig_encoding}). 

\subsection{Encoding Block}\label{encoding_block}

Previous works, which do not use additional NNs for pre- or postprocessing, generally design the encoding block of the VQC such that each feature of the observation space is encoded into one qubit using angle embedding. This strategy limits the size of the VQC to the observation space size of the task at hand, limiting the potential of VQCs. This problem is accompanied by the fact that an increase of the number of layers has previously been shown to improve training performance only until a certain threshold  \cite{skolik2022quantumGym}. 

To overcome this issue we propose a novel data encoding approach in order to increase the number of exploitable qubits: Instead of encoding each feature of the state $s$ using angle embedding only once, we stack $s$ such that each feature is encoded several times. An illustration of this \textit{stacked} VQC is shown in Fig. \ref{fig:vqc_stacked}. This architecture enables VQC-based QRL agents to scale beyond the previous VQC sizes, potentially enhancing their training capabilities due to a higher amount of trainable parameters without the need of additional layers. 

\begin{figure*}
    \centering 
\begin{subfigure}{0.32\textwidth}
  \includegraphics[width=\linewidth]{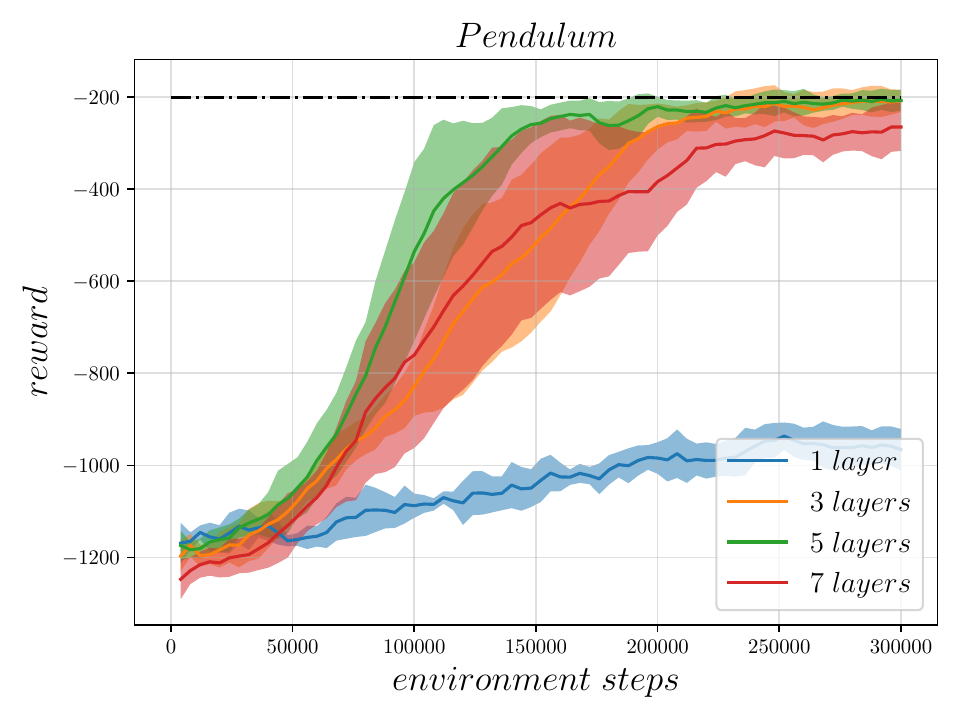}
  \caption{3 qubits, $\hat s_i \cdot \lambda_i$ encoding, $M_1$ readout}
  \label{fig_single}
\end{subfigure} 
\medskip
\begin{subfigure}{0.32\textwidth}
  \includegraphics[width=\linewidth]{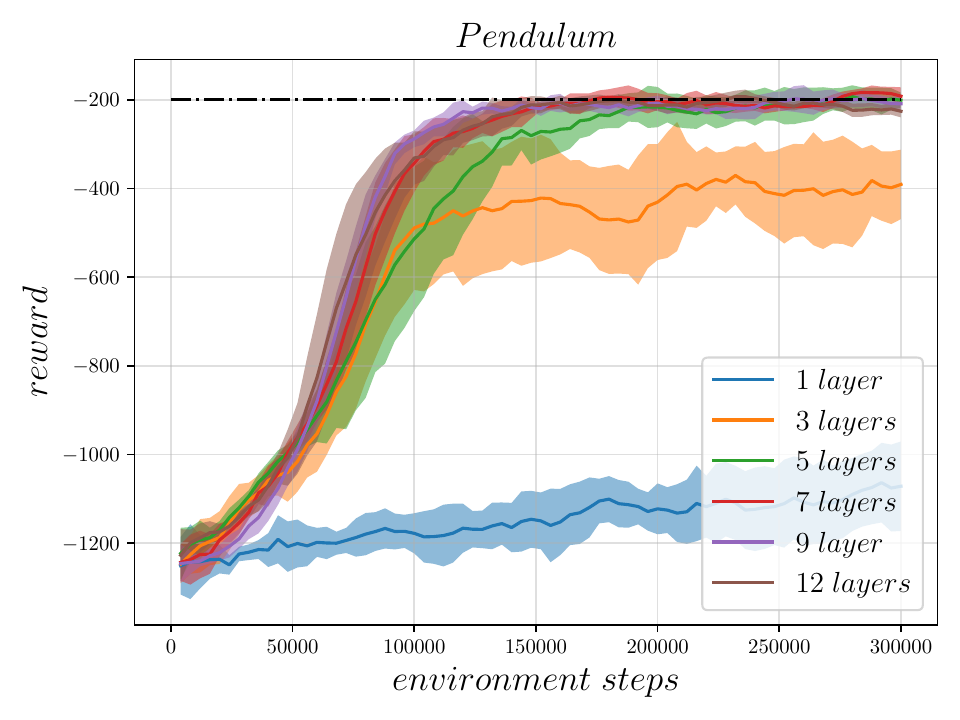}
  \caption{6 qubits, $\hat s_i \cdot \lambda_i$ encoding, $M_1$ readout}
  \label{fig_double}
\end{subfigure}
\medskip
\begin{subfigure}{0.32\textwidth}
  \includegraphics[width=\linewidth]{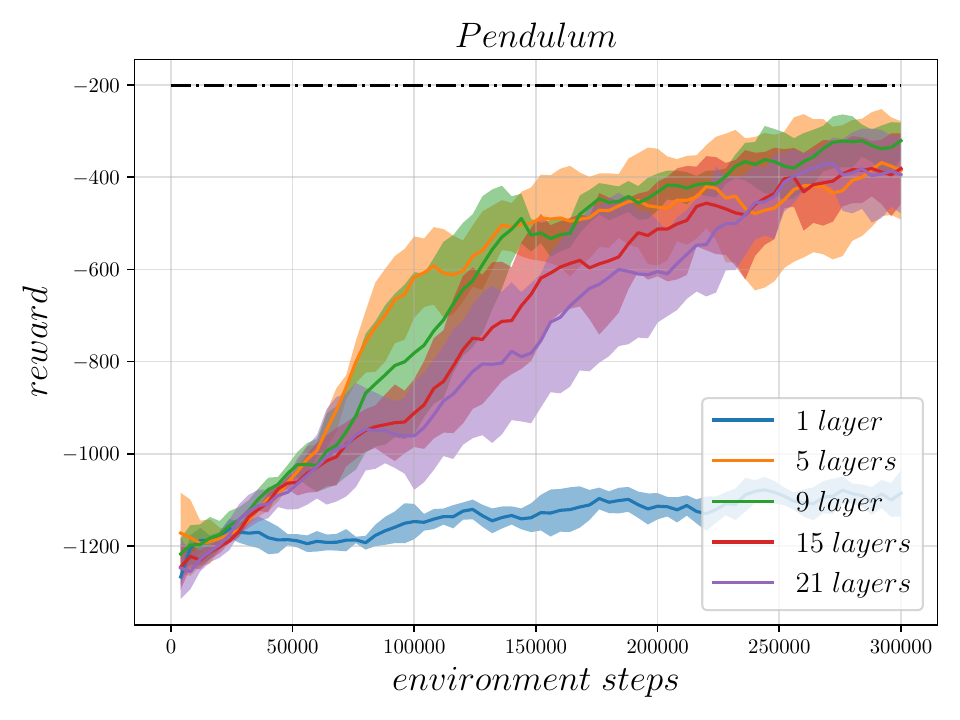}
  \centering
  \caption{9 qubits, $\hat s_i \cdot \lambda_i$ encoding, $M_1$ readout}
  \label{fig_triple}
\end{subfigure}

\caption{Comparison of different VQC encoding block sizes: In all runs the same angle embedding ($\hat s_i \cdot \lambda_i$) and readout ($M_1$, ref. Tab. \ref{measurement_table}) is used. The state features $s_i$ are encoded one, two or three times in Fig. \ref{fig_single}, \ref{fig_double} and \ref{fig_triple} respectively as depicted in Fig. \ref{fig:vqc_stacked}. Each solid line represents the mean of five seeds, the shaded area indicates the standard deviation.}
\label{fig_encoding_block}
\end{figure*}

\subsection{Observables and Postprocessing}

The choice of observables and postprocessing steps, jointly referred to in this work as readout configuration, has previously been shown to be crucial for the performance of the agent on discrete learning tasks \cite{meyer2023quantum}. Therefore, the choice of the observables $O_{\mu_i}$ and $O_{\sigma_i}$ for the actor is investigated in this work. 

For the actor VQC, we compare single qubit observables for the mean $O_{\mu_i} = Z_i$ and variance $O_{\sigma_i} = Z_{i+1}$, where $Z_i$ are Pauli-Z operators on the respective qubit. We compare these single-qubit observables to multi-qubit observables, as well as to a combination of the two approaches (ref. Tab. \ref{measurement_table}). Since the expectation value of the unscaled observables $O_{\mu_i}$ and $O_{\sigma_i}$ lie in $ [-1, 1] $, while the continuous action space of a given environment can potentially lie in $(-\infty, \infty)$, we use one trainable parameter $w_i$ for each observable as postprocessing step. Previous work has already analyzed the impact of non-trainable scaling parameters for Q-learning \cite{skolik2022quantumGym}, so we will not investigate this design choice here. 

For the critic, the observable $O_{v_i}$ is either a single Pauli-Z operator on the first qubit, the sum of single Pauli-Z operators on all qubits or a multi qubit measurement on all qubits. As postprocessing step we introduce for each respective expectation value a trainable scaling parameter $w_i$. In the following we investigate the impact of different readout configurations $M$ - with varying observables and postprocessing steps - on the training performance of the QRL agent.

\section{\uppercase{Numerical Results}}

In this section we analyze the influence of the VQC design choices on two CAS environments with different observation space sizes, action space sizes and difficulties: The Pendulum-v1 environment with observation space of size three and one continuous action and the LunarLander-v2 environment with observation space of size eight and two continuous actions. On Pendulum-v1, we benchmark VQCs with 3, 6 and 9 qubits and evaluate different design choices. On LunarLander-v2, we select the 6 most informative features of the 8 features of the observation space. This is because for the used VQC architecture, the variance of the expectation values of the observables starts to vanish quickly, hindering training already at eight qubits. For details on the environments we refer to \cite{brockman2016openai} and for hyperparameters to Tab. \ref{hyper_table} in the Appendix. All calculations were performed using statevector simulators.

\subsection{Angle Embedding}

In Fig. \ref{fig_encoding} various angle embedding strategies are evaluated on the two environments. In Figs. \ref{pendulum_without_norm} and \ref{lunarlander_without_norm}, the state features $s_i$ are encoded into two rotation gates $R_y$ and $R_z$ using $arctan$ or $sigmoid$ functions and scaling parameters $\lambda_i$. None of these encodings enable the VQC-based QRL agents to solve the environments. On both benchmarks the best performing agents utilize a $sigmoid(s_i \cdot \lambda_i)$ encoding, but nevertheless fail to reach the target rewards.

In classical RL, input states are generally normalized to the interval of $[-1, 1]$ in order to enhance training performance. Following this idea of previous normalization of states, in Fig. \ref{pendulum_with_norm} and \ref{lunarlander_with_norm} $s_i$ is previously normalized to the interval of $[-\pi/2, \pi/2]$ to $\hat s_i$. In Fig. \ref{pendulum_with_norm} it can be seen that the encoding without any nonlinear function scaling outperforms all other encodings, while in Fig. \ref{lunarlander_with_norm} the encodings using $arctan$ functions perform similarly well. 

While the use of nonlinear functions such as $arctan$ and $sigmoid$ is widely spread across literature, we show that they do not enhance training performance, but instead can even lead to poorer results. Instead, simple normalization as utilized in classical RL, combined with a trainable parameter for each input feature shows the best performance across runs. Hence, in the further comparison, the encoding $\hat s_i \cdot \lambda_i$ is used and referred to as basic encoding. 

\begin{figure*}[ht!]
    \centering 
\begin{subfigure}{0.33\textwidth}
  \includegraphics[width=\linewidth]{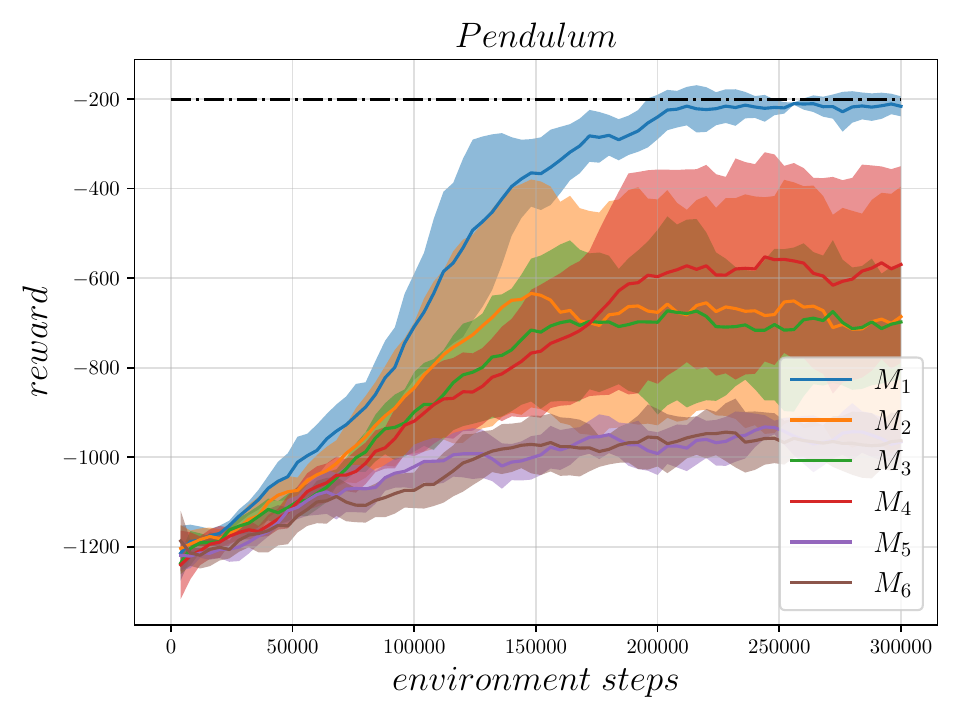}
  \caption{3 qubits, 5 layers, $\hat s_i \cdot \lambda_i$ encoding}
  \label{pendulum_measurement_3}
\end{subfigure}\hfil 
\medskip
\begin{subfigure}{0.33\textwidth}
  \includegraphics[width=\linewidth]{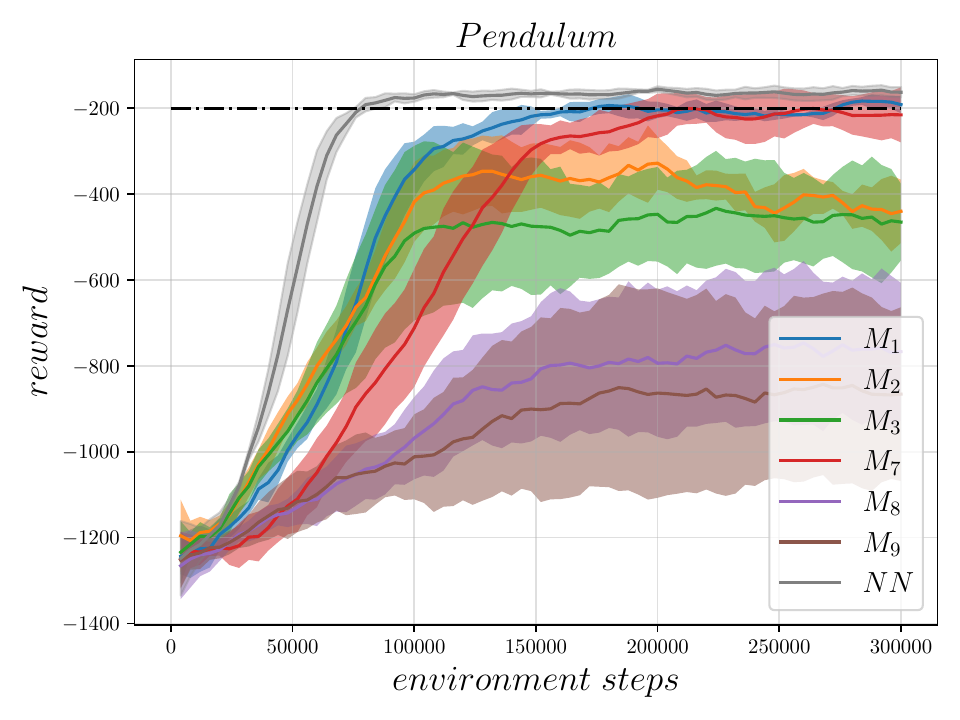}
  \caption{6 qubits, 7 layers, $\hat s_i \cdot \lambda_i$ encoding}
  \label{pendulum_measurement_6}
\end{subfigure}
\medskip
\begin{subfigure}{0.33\textwidth}
  \includegraphics[width=\linewidth]{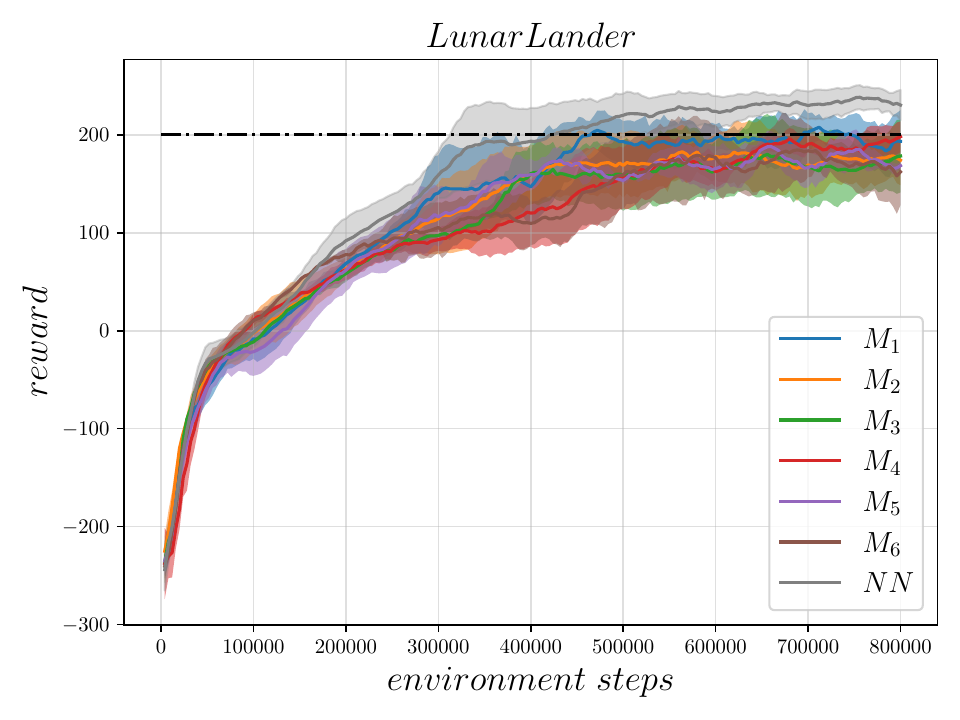}
  \caption{6 qubits, 7 layers, $\hat s_i \cdot \lambda_i$ encoding}
  \label{lunarlander_measurement_6}
\end{subfigure}
\caption{Comparison of readout configurations (ref. Tab. \ref{measurement_table}) and benchmarking against best performing NNs: In Fig. \ref{pendulum_measurement_6} and \ref{lunarlander_measurement_6} the black lines indicate the best performing classical NN with two hidden layers with 64 neurons each and ReLU activation functions based on the hyperparameter search as described in Tab. \ref{hyper_table}. Each solid line represents the mean of five seeds, the shaded area indicates the standard deviation.}
\label{fig_measurement}
\end{figure*}

\subsection{Encoding Block}

Previous works, which do not use additional NNs for pre- or postprocessing, have always used a VQC of the same size as the observation space of the classical environment. In order to evaluate the usage of different sized encoding blocks, in Figs. \ref{fig_encoding_block} we compare the training performance on Pendulum-v1 with a 3, 6 and 9 qubit VQC, where each feature of the state $s$ is encoded one, two and three times respectively (ref. Fig. \ref{fig:vqc_stacked}). All VQCs fail to train with a single layer, even though the amount of trainable parameters in the variational block increases from 6 to 12 and 18. As data reuploading is introduced by using more than one layer, training performance is improved across all architectures. But simply introducing more parameters by adding more layers has previously been shown to only improve performance until a certain threshold \cite{skolik2022quantumGym}. This can be seen in Fig. \ref{fig_single}, where performance reaches a peak at five layers. Skolik et al. suggest that this behaviour occurs because at a certain number of layers, training begins to be hindered due to overparameterization of the VQC. Interestingly, the amount of layers required for successful training increases with the amount of qubits: For larger VQCs with six qubits best performance is observed for nine layers (Fig. \ref{fig_double}), while the nine qubit VQC fails to solve the environment in the given time frame (Fig. \ref{fig_triple}). Therefore, overparameterization can not be interpreted as an absolute number of trainable parameters, but rather depends on the number of qubits used: Our results indicate that greater qubit numbers also require greater numbers of trainable parameters. On the Pendulum-v1 environment, the best performing agent across all runs has a VQC with six qubits and nine layers, having more than two times the amount of trainable parameters as the best performing three qubit VQC with five layers.

It has been previously shown that for small VQCs the amount of trainable parameters required for successful training is lower than for classical NNs \cite{druagan2022quantum}. Our findings suggest that this phenomenon is restricted to small VQCs and does not apply to larger VQCs. Moreover, the vanishing gradients start to hinder training already at nine qubits (ref. Fig. \ref{fig_triple}).

\subsection{Observables and Postprocessing}

In Fig. \ref{fig_measurement} the influence of different choices of observables and postprocessings for actor and critic is shown. The different readout configurations are listed in Tab. \ref{measurement_table}. On the Pendulum-v1 environment the choice of observable is crucial for the success of training (Fig. \ref{pendulum_measurement_3}). The $M_1$ readout configuration is the only configuration which leads to successful training with a three qubit VQC. Also in Fig. \ref{pendulum_measurement_6}, the $M_1$ readout performs best for the \textit{stacked} VQC, followed by $M_7$, the only other configuration leading to successful training. Finally, in Fig. \ref{lunarlander_measurement_6} different readout configurations for LunarLander-v2 are shown. Here no clear trend can be observed.

Our results show that observables and postprocessing steps can be crucial for training performance in some cases, while in others barely influence the performance of the agents. Only the $M_1$  readout configuration has no negative influence across all experiments. 

\subsection{Benchmark against Classical Agents}

Finally we perform an extended hyperparameter search for the classical RL agents and benchmark the best performing classical RL agents against the QRL agents in Figs. \ref{pendulum_measurement_6} and \ref{lunarlander_measurement_6}. We evaluated 117 different classical agents for Pendulum-v1 and 36 for LunarLander-v2 (ref. Tab. \ref{hyper_table}). On both benchmarks the best performing NNs have two hidden layers with 64 neurons each and ReLU activation functions, resulting in 4416 and 4736 trainable parameters on the two benchmarks, while the QRL agents have 176 and 178 trainable parameters. On one hand, the performance gap could be explained by the difference of the number of trainable parameters. On the other hand, it remains to be shown if the beneficial properties such as better trainability \cite{abbas2021power} and generalization  \cite{banchi2021generalization} also hold for larger VQCs with comparable amounts of trainable parameters.

\section{\uppercase{Conclusions}}
\label{sec:conclusion}

In this work we showed how to construct a quantum reinforcement learning agent for classical environments with continuous action spaces based on a hybrid quantum-classical algorithm that employs variational quantum circuits as function approximators. Our approach does not require any additional classical neural network layers as pre- or postprocessing steps. Instead, only trainable scaling parameters are required in order to adapt the output of the variational quantum circuit to the size of the continuous action space. 

Additionally, we investigated several variational quantum circuit design choices with respect to their influence on training performance. While nonlinear functions such as $arctan$ have been widely used throughout quantum reinforcement learning literature for angle embedding, we show in our experiments that such functions actually hinder training performance. Instead, normalization - in combination with trainable scaling parameters - yields the best training results. 

The number of qubits of previous designs of variational quantum circuits was limited to the size of the observation space due to angle embedding. We proposed a new encoding block architecture - \textit{stacked} VQC - which allows the utilization of additional qubits, resulting in improved training performance. It has been previously shown that an increase of the number of layers improves training performance only until a threshold \cite{skolik2022quantumGym}. We reveal a similar trend: an increase of the number of qubits substantially improves training performance, but also only until a certain limit. Our work indicates that current VQC architectures therefore are limited both in the number of layers, as well as in the amount of qubits, and thus dictate both the depth and the width of the circuit, respectively. While we investigated and enhanced current variational quantum circuit design choices, future work should aim to further improve upon these results as well as explore novel circuit architectures in order to bridge the performance gap between QRL and RL.


\section*{\uppercase{Acknowledgements}}
The research is part of the Munich Quantum Valley, which is supported by the Bavarian state government with funds from the Hightech Agenda Bayern Plus.

\bibliographystyle{apalike}
{\small
\bibliography{bibliography}}

\newpage

\onecolumn
\section*{\uppercase{Appendix}}

\begin{table*}[ht]
\centering
\begin{tabular}{p{0.5cm} | c || c || c} 
 &  $O_{\mu_i}$ &  $O_{\sigma_i}$ &  $O_{v}$ \\
 \hline
  \hline
 $M_1$ & $Z_i  w_{\mu_i}$ & $ Z_{i+1}  w_{\sigma_i}$    & $\sum (Z_j \cdot w_{v_j})$ \\
 $M_2$ & $Z_i  w_{\mu_i}$ & $ Z_{i+1}  w_{\sigma_i}$    &  $Z_0 \cdot w_{v_0}$    \\
 $M_3$ & $Z_i  w_{\mu_i}$ & $ Z_{i+1}  w_{\sigma_i}$    & $\prod (Z_j) \cdot w_{v_0}$ \\
 $M_4$ & $Z_i Z_{i+1} w_{\mu_i}$ &  $Z_{i+2} w_{\sigma_i}$    & $\sum (Z_j \cdot w_{v_j})$ \\
 $M_5$ & $Z_i Z_{i+1} w_{\mu_i}$ &  $Z_{i+2} w_{\sigma_i}$    &  $Z_0 \cdot w_{v_0}$   \\
 $M_6$ & $Z_i Z_{i+1} w_{\mu_i}$ &  $Z_{i+2} w_{\sigma_i}$    & $\prod (Z_j) \cdot w_{v_0}$\\
 $M_7$ & $Z_i Z_{i+1}Z_{i+2} w_{\mu_i}$ &  $Z_{i+3} Z_{i+4}Z_{i+5} w_{\sigma_i}$    & $\sum (Z_j \cdot w_{v_j})$ \\
 $M_8$ & $Z_i Z_{i+1}Z_{i+2} w_{\mu_i}$ &  $Z_{i+3} Z_{i+4}Z_{i+5} w_{\sigma_i}$   &  $Z_0 \cdot w_{v_0}$   \\
 $M_9$ & $Z_i Z_{i+1}Z_{i+2} w_{\mu_i}$ &  $Z_{i+3} Z_{i+4}Z_{i+5} w_{\sigma_i}$    & $\prod (Z_j) \cdot w_{v_0}$\\
 \hline
 \hline
\end{tabular}
\caption{Table of all readout configurations $M_1$ to $M_7$ in Fig \ref{fig_measurement}, where $i$ is the index of the action $a_i$ and $j$ is the index of the qubit.}
\label{measurement_table}
\end{table*}

\begin{figure*}[ht]
        \centering
        \includegraphics[width=0.8\textwidth]{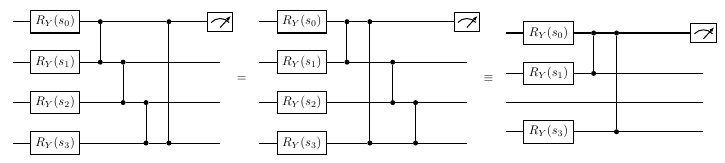}
        \caption{
        The reason why $CZ$ gates lead to parameters not influencing the measurement output is that $Z$ gates are diagonal, thus they and their controlled variants commute (as opposed to CNOT gates): $I \otimes \ket{0}\bra{0} + Z \otimes \ket{1}\bra{1} = \ket{0}\bra{0} \otimes I + \ket{1}\bra{1} \otimes Z$, meaning that the control wire and the acting wire may be swapped at will.To illustrate this behaviour, a simple example is depicted above: We start with a four qubit VQC with chain entangling, where just one rotation gate is used to encode the state features $s_i$. We only measure the first qubit, e.g. for the mean $\mu_i$ of a continuous action $a_i$ (Eq. \ref{gaussian_policy_mean}). In the first step, we use the fact that any two controlled $Z$ gates commute. We reformulate the circuit such that it becomes apparent that the gates which do not directly act on the measurement wire cannot influence the measurement outcome. In step two, we have thus omitted these gates as well as the $R_y$ gate which encodes feature $s_2$. This feature no longer influences the action $a_i$ due to the locality of the $CZ$ entangling gates. This behaviour can evidently be harmful for training, so CNOT gates are used in this work.
        }
        \label{triangular:fig:explanation}
\end{figure*}

\begin{table*}[ht!]
\begin{tabular}{p{0.20\textwidth}||p{0.41\textwidth}||p{0.30\textwidth}}
Parameter & Pendulum-v1 & LunarLander-v2  \\
\hline
\hline
Learning Rate          & 0.01, 0.005, 0.001        &   0.001, 0.0005, 0.0001     \\
Activation Functions    & ReLU, leaky ReLU, tanh    &   ReLU, leaky ReLU, tanh  \\
\text{[$n_1$, $n_2$]} &  [8], [16], [32], [64], [8,8], [16,16], [32,32], [64,64], [8,16], [8,32], [16,32], [16,64], [32,64] & [8,8], [16,16], [32,32], [64,64]  \\
\hline
\end{tabular}
\caption{Table of the hyperparameters used for the grid search with a total of 117 configurations for Pendulum-v1 and 36 for LunarLander-v2. The number of neurons in the first hidden layer of the actor and the critic neural networks is noted as $n_1$, while the number of neurons in the second hidden layer (if applicable) is $n_2$. Batch size and other unmentioned parameters are kept constant and of the same values as in Tab. \ref{hyper_table}.}
\label{grid_table}
\end{table*}

\begin{table*}[ht!]
\begin{tabular}{ p{0.32\textwidth}||p{0.32\textwidth}||p{0.32\textwidth}}
Parameter & Pendulum-v1 & LunarLander-v2  \\
 \hline
  \hline
 Learning Rate for $\theta$,  $\phi$  & 0.001    & 0.0005   \\
 Learning Rate for $\lambda$ & 0.001    & 0.0005    \\
 Learning Rate for $w$       & 0.01     & 0.01  \\
 Learning Rate Scheduler  & -        & start: 300.000 steps \\
 & & decay factor: 0.97 \\
 Initialization $\theta$,  $\phi$     & $\mathcal{N}(0, 0.1)$ & $\mathcal{N}(0, 0.1)$\\
 Training Batch Size        & 4000 & 4000  \\
 SGD mini Batch Size       & 64   & 64 \\
 Number of SGD iterations     & 10   & 10 \\
 GAE factor $\lambda$ & 0.1  & 0.99 \\
 Discount factor $\gamma$  & 0.99 & 0.99 \\
 \hline
\end{tabular}
\caption{Table of the hyperparameter values for our hybrid quantum-classical variational algorithm for all experiments. The first three rows present the values of the learning rates, set individually for the variational parameters $\theta$ and $\phi$, the input scaling parameters $\lambda$ and the output scaling parameters $w$.}
\label{hyper_table}
\end{table*}

\end{document}